# Supernova Impostors and other Gap Transients

*Besides supernovae, few astrophysical processes can release close to $10^{51}$ erg of energy. A growing number of stellar outbursts are now recognised to have energy releases matching those of faint supernovae. These transients can be triggered by a variety of mechanisms, and their discrimination is sometimes a tricky issue.*

The death of a massive star in a supernova (SN) explosion represents the sudden release of $\sim 10^{51}$ erg (1 foe) as radiation and kinetic energy. Ongoing all-sky transient surveys have now begun to find significant numbers of less energetic transients that are fainter and have lower kinetic energy than normal SNe, but are more luminous than classical novae (Fig. 1). These transients are often called "gap transients" (ref. [1]), and pose a fundamental question: are they also associated with terminal explosions of massive stars, or do they represent new physical classes of transient phenomena?

Among gap transients, stellar mergers can release around $10^{49}$ erg. Alternatively, some massive stars can also undergo non-terminal eruptions and outbursts (and these are termed "supernova impostors"). Some supernova impostors appear to presage the explosion of a massive star as a genuine core-collapse supernova — in such cases the supernova impostor must be associated with an instability in the latter stages of stellar evolution. Other supernova impostors appear to be eruptions long before the star will undergo core-collapse, as exemplified by the Great Eruption of Eta Carinae in the middle of the 19th century.

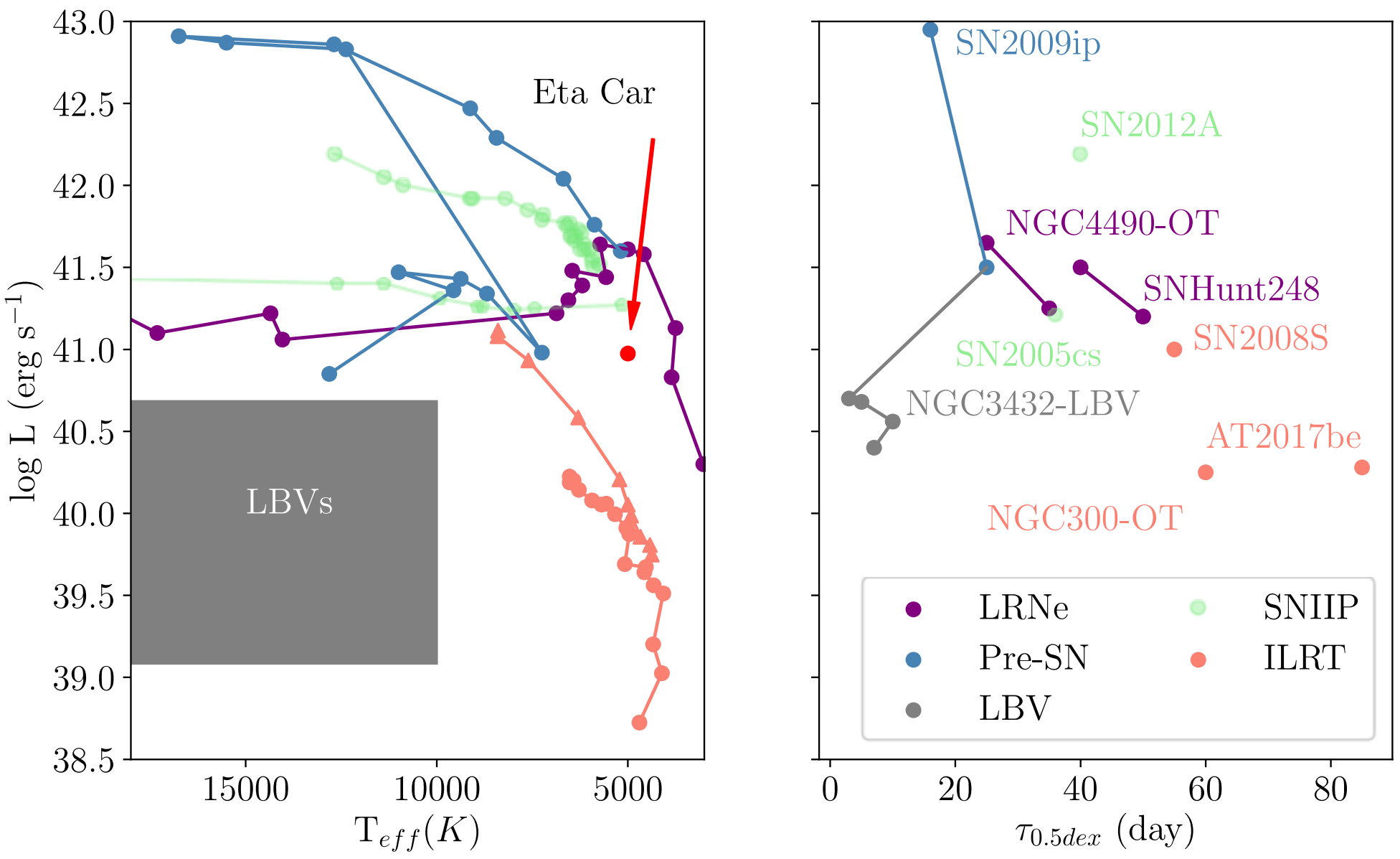

Fig. 1| **Summary of the main properties of gap transients.** Luminosity versus effective temperature ($T_{\rm eff}$) (left) and timescale for a luminosity decline of 0.5 dex (right) diagrams for the gap transients discussed in this article. For comparison, the reference parameters for two type II-P SNe (2005cs and 2012A) are also shown.

Distinguishing between the various classes of non-terminal eruptions and outburst, and faint hydrogen-rich core-collapse supernovae remains a key challenge in transient astronomy. Here we propose a physical characterization of the different subtypes of gap transients on the basis of some common observational properties (see, also Fig. 1).

- **Faint core-collapse supernovae.** Some faint "gap transients" can actually be genuine supernovae (making them "SN impostor impostors"). In particular, the class of Intermediate-Luminosity Red Transients (ILRTs; e.g. ref. [2]) have attracted attention as possible examples of weak electron-capture supernovae. The prototypical SN 2008S (ref. [3]) had a slow rise to maximum light (with $M_R$≈−15 mag), followed by a linear decline lasting about 4 months. Other objects peaked at −12 to −14 mag, and have shown a post-peak plateau followed by a rapid decline 3−4 months later. When observed at late phases, the light curves settle onto a slower decline rate, close to that expected from the decay of $^{56}$Co. These light curves resemble those of very faint Type II-L or II-P SNe. The spectra of ILRTs are initially blue, and become redder with time. They always show Balmer lines in emission, along with prominent Ca II features (in particular, the [Ca II] λλ7291, 7323 doublet is always detected, see Fig. 2). Together, these properties have been taken as suggestive of a weak SN explosion inside a dense cocoon of circumstellar material (CSM).

The progenitor stars of some ILRTs have been detected in quiescence in mid-infrared Spitzer archive images. In contrast, deep optical and near-IR images (including HST data) obtained prior to outburst do not show any source at the ILRT position. While most agree that the progenitors are 8−15 $M_\odot$ stars embedded in dusty cocoons, the nature of the transients is somewhat controversial. Along with the faint SN scenario, the ejection of a common envelope due to binary interaction, the eruptive formation of a relatively massive white dwarf or a super-Eddington event of a moderate-mass star were proposed for ILRT outbursts (e.g. ref. [4]). However, more recent Spitzer observations of sites of ILRTs a few years after the outburst show that the residual flux at the transient position is much fainter than the quiescent progenitor, hence supporting terminal SN explosions (ref. [5]). If ILRTs are core-collapse SNe, they necessarily eject very modest $^{56}$Ni masses ($10^{-3}$ to $10^{-4}$ $M_\odot$). This, and the clear evidence of circumstellar gas and dust, point towards an electron-capture SN from super-AGB stars for all ILRTs (e.g., ref. [6]).

- **Mergers.** Often referred to as "Luminous Red Novae" (LRNe), these transients display a slow rise over months to years, followed by a distinctive double peaked lightcurve (see Fig. 2, top panel). During the first peak, optical spectra appear blue and show prominent emission lines of H and Fe II. During the second peak, we observe a major metamorphosis in the spectrum, which becomes similar to a late G to K-type star, with a forest of narrow metal lines in absorption (with $v$~100 km s$^{-1}$), and much weaker Balmer lines. Finally, at late phases, during the fast, late luminosity decline, the spectrum transitions to that of an M-type star, with Hα becoming prominent again, while the optical spectrum, now very red, shows strong molecular absorption bands (mostly TiO and VO; Fig. 2, bottom panel).

  The strongest evidence that such transients come from mergers in binary systems comes from studies of the Galactic transient V1309 Sco (ref. [7,8]). Crucially, periodic modulations in brightness were observed superimposed on a trend of slowly rising luminosity. The period seen in photometry slowly decreased over time, consistent with two in-spiraling stars in a binary system. The subsequent four-magnitude lightcurve rise in about 6 months and the disappearance of the photometric periodicity were interpreted as being due to the ejection of a common envelope. A sudden brightening by further 4 magnitudes over a few days and the following double-peaked light curve were due to a violent gas ejection as a consequence of the stellar coalescence, with the gas outflow later interacting with the previously ejected common envelope.
  LRNe span a wide range of absolute magnitudes, from −4 to −15, with frequent faint events being produced by low mass mergers, while rare luminous events are produced by the coalescence of stars in massive binaries (a few tens solar masses; ref. [9,10]).

- **Supernova impostors.** Supernova impostors (ref. [11,12]) are non-terminal eruptions from massive stars, and are often associated with Luminous Blue Variables (LBVs). The LBV phase is a short-duration stage of stellar life, during which a very massive star (≥40$M_\odot$) becomes an evolved hypergiant with luminosity of several $10^5$ $L_\odot$ (ref. [13]). LBVs may undergo long-lasting giant eruptions, as observed during past centuries in the Milky Way: Eta Car erupted during the second half of the 19th century later producing the spectacular Homunculus nebula, while P Cygni experienced repeated outbursts around four centuries ago. In both cases, several to several tens of solar masses of material were ejected from the star.

  Although very rare, giant eruptions can be also observed in other galaxies. Well-known cases include the LBV in NGC 3432 (also known as SN 2000ch; Fig. 2) which produced multiple outbursts up to $M_R$~−14 mag in the last 3 decades (ref. [14]). Occasionally, long-term smooth variability or isolated outbursts are also observed in massive hypergiants (ref. [15,16]). Spectroscopically, such events are typified by blue continua and narrow (~$10^2$ to $10^3$ km s$^{-1}$) emission lines. The emission lines result from the photoionization of the circumstellar material lost by the star. The mechanism behind such major mass-loss events is still unclear. Close stellar encounters in close binary systems is a viable explanation for some Giant Eruptions, while super-Eddington continuum-driven winds have been also invoked for single stars (ref. [17]), although what leads a star to suddenly exceed the Eddington luminosity/mass limit by a large amount is still a puzzle.

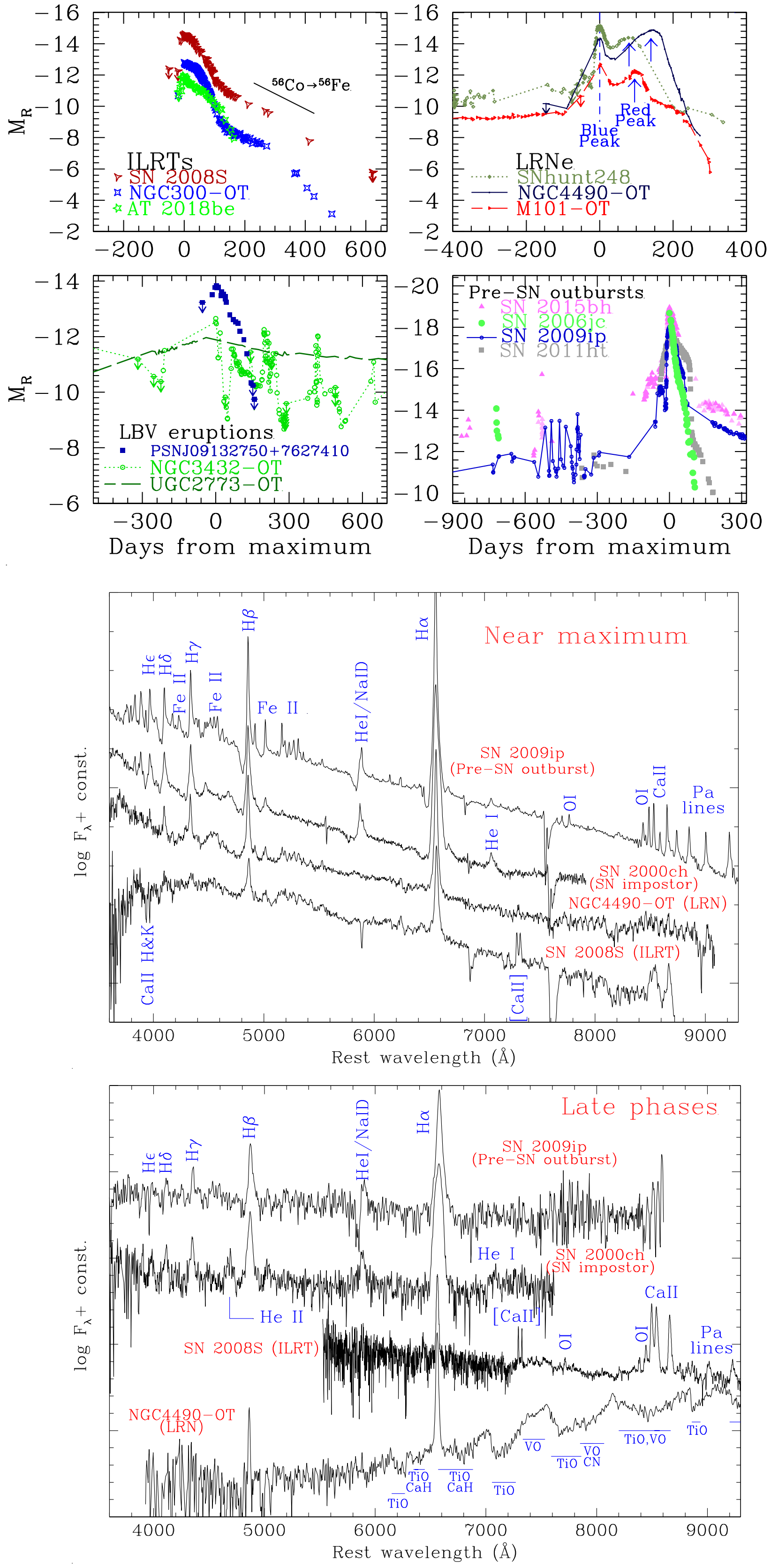

Fig. 2 | **The diversity of gap transients.** Top panel: Representative light curves of gap transients subtypes. Middle panel: spectral sample of gap transients near the maximum light, including a pre-SN outburst (SN 2009ip), a SN impostor in outburst (SN 2000ch), a LRN (NGC4490-OT; during the first, blue peak) and an ILRT (SN~2008S). Bottom panel: spectra of the above sample, but obtained several months after the outburst maximum. The spectra of SN 2009ip were obtained over two years before the putative SN explosion. Data are from ref. [3,10,14,15,16,19,21,22,26,27,28,29,30,36,37,38,39,40,41,42,43,44,45,46,47], along with unpublished data from our team.

- **Pre-supernova outbursts.** In some cases SN impostors likely heralded the death of a star (e.g. ref. [18, 19]; Fig. 2 top). The SNe that follow these outbursts are usually classed as type IIn (ref. [20]), with spectra dominated by narrow H emission lines (FWHM ~$10^3$ km $s^{-1}$). These lines are indicative of the interaction of fast SN ejecta with slower moving H-rich CSM ejected immediately prior to the SN explosion.
  However, impostors are not only observed before type IIn SN explosions, or solely from LBVs. A pre-SN outburst of $M_R \approx -14$ mag, in fact, was observed by the amateur astronomer K. Itagaki two years before the explosion of SN 2006jc (ref. [21, 22]; Fig. 2, top panel). The SN spectrum revealed narrow emission lines of He and no H, hence it was classified as Type Ibn SN (ref. [23]). The outburst was likely a dramatic mass-loss event from a Wolf-Rayet (WR) star that produced a He-rich cocoon. The WR star later exploded as a stripped-envelope SN, as confirmed by a post-death inspection of Hubble Space Telescope images (ref. [24]).

  At present, stellar evolutionary models do not predict outbursts immediately prior to a SN explosion. However, the latter stages of stellar evolution are notoriously difficult to simulate, and recent work has suggested that violent convection and instabilities associated with late nuclear burning stages could lead to eruptions. With a better understanding of what drives pre-SN outbursts, and how common they are, one may even envisage being able to predict stars that are just about to undergo core-collapse.

Although the nature of gap transients is not yet fully understood, some firm markers have been posed over the last decade: all of them are linked to moderate to high mass stars; these stars are enshrouded by H-rich cocoons, and they show signatures of ejecta-CSM interaction; and in some cases, the outbursts are followed by a much brighter event which is possibly the terminal SN explosion.

The object that perhaps most clearly illustrates these characteristics is SN 2009ip, where a very massive star of over 40 $M_\odot$ (ref. [25]) was observed to repeatedly erupt for at least 3 years (hence it was formally a SN impostor; ref. [26, 27]). Then, in summer 2012, the object went into outburst again, with a light curve initially reaching a peak of −15.5 mag, before brightening further to −19 three weeks later (ref. [28, 29, 30]). Recently, a number of transients with properties similar to those of SN 2009ip have been later discovered (e.g., ref. [31], and references therein). Several scenarios were proposed to explain the sequence of events observed in SN 2009ip-like events, including a giant outburst followed by the SN explosion; a faint SN explosion followed by major ejecta-CSM interaction; a giant outburst due to extreme binary interaction or pulsational pair-instability in an LBV progenitor followed by the collision between massive shells. Whether the progenitor of SN 2009ip really exploded as a SN or not has still not been conclusively settled. However, the late-time light curve evolution of SN 2009ip with an almost linear decline to very faint magnitudes and no further outbursts seems to support a terminal Type IIn SN explosion.

Ongoing all-sky surveys have significantly increased the number of known gap transients, and many more will be discovered in the Large Synoptic Survey Telescope era. Although significant progress has been made in discriminating between the different types of gap transients (Fig. 2), the triggering mechanisms and the fate of their progenitor stars remain debated. Deep imaging of the explosion sites a few years after the outbursts with space facilities like HST and Spitzer, or future ground-based telescopes with adaptive optics (such as the Extremely Large Telescope) can reveal whether the progenitor star of a gap transient has disappeared as a consequence of a dim SN explosion or survived the outburst (e.g., ref. [5,32]). For the survivors, the comparison of their magnitude and colour information with theoretical evolutionary tracks, will provide crucial parameters such as zero-age main sequence mass, final mass and eventually binary signatures. Spectroscopy at very late phases (with 10-m class telescopes) or multi-domain monitoring, are also powerful tools to disentangle alternative scenarios for controversial events (ref. [33,34,35]). All of this will enable us to address the remaining uncertainty on the nature of gap transients.

Andrea Pastorello[1*] and Morgan Fraser[2]
[1] *INAF – Osservatorio Astronomico di Padova, Padova, Italy.*
[2] *School of Physics, O'Brien Centre for Science North, University College Dublin, Dublin, Ireland.*
* *e-mail:* andrea.pastorello@inaf.it

## Acknowledgements

M.F. is supported by a Royal Society – Science Foundation Ireland University Research Fellowship.
A.P. and M.F. thank their many students, collaborators and colleagues with whom they have had the pleasure of working with on this topic. Without their ongoing collaboration, this article could not have been written.